\documentclass[5p,twocolumn]{elsarticle}
\usepackage{amssymb}
\usepackage{bm}
\usepackage{graphicx}%\
\biboptions{sort&compress}
\begin{document}
%\draft
\title
{Transport properties of $\nu=1$ quantum Hall bilayers.
Phenomenological description.}

\author[dvf]{D.V.~Fil\corref{cor1}}
\ead{fil@isc.kharkov.ua}
\author[sis]{S.I.~Shevchenko}
\cortext[cor1]{Corresponding author}
%\email{fil@ilt.kharkov.ua}
\address[dvf]{%
Institute for Single Crystals, National Academy of Sciences of
Ukraine, Lenin av. 60, Kharkov 61001, Ukraine}
\address[sis]{B.~Verkin
Institute for Low Temperature Physics and Engineering, National
Academy of Sciences of Ukraine, Lenin av. 47 Kharkov 61103,
Ukraine}

\begin{abstract}
 We propose a phenomenological model
that describes counterflow and drag experiments with quantum Hall
bilayers in a $\nu_T=1$ state. We consider the system consisting
of statistically distributed areas with local total filling
factors $\nu_{T1}>1$ and $\nu_{T2}<1$. The excess or deficit of
electrons  in a given area results in an appearance of vortex
excitations. The vortices in quantum Hall bilayers are charged.
They are responsible for a decay of the exciton supercurrent, and,
at the same time, contribute to the conductivity directly. The
experimental temperature dependence of the counterflow and drive
resistivities is described under accounting viscous forces applied
to vortices that are the exponentially increase functions of the
inverse temperature. The presence of defect areas where the
interlayer phase coherence is destroyed completely can result in
an essential negative longitudinal drag resistivity as well as in
a counterflow Hall resistivity.
\end{abstract}
%\pacs{}

\maketitle

\section{Introduction}
Theoretical prediction on superfluidity of bound electron-hole
pairs in electron-hole bilayers \cite{1, 1-1,  1-3,  1-5} and,
especially, in quantum Hall bilayers
\cite{fer,3,4,5,d1,d2,d5,d6,5-1} has stimulated experimental study
of the counterflow transport in such systems. In these experiments
\cite{8,8-1,9,10,11,12} electrical current
 is passed through one layer in a given direction and returned to
the source through the other layer in the opposite direction. The
counterflow currents in the adjacent layers can be provided by
excitons that consist of an electron belonging to one layer and a
hole belonging to the other layer. One can expect that at low
temperature the gas of such excitons becomes superfluid and the
counterflow current may flow without dissipation.

In the counterflow experiments  \cite{8,8-1,9,10,11,12} bilayer
quantum Hall systems with the total filling factor $\nu_T=1$ are
used.  At such a filling the number of electrons in one layer
coincides with the number of empty quantum states (holes) in the
lowest Landau level  in the other layer. At rather small
interlayer distances \cite{5-2} the ground state of such a system
is the BCS-like state with electron-hole pairing. The pairing is
caused by the Coulomb attraction. The BCS-like state can also be
considered as  a superfluid state of interlayer excitons.

In spite of expectation to reach at low temperature  zero
electrical resistance, counterflow experiments in quantum Hall
bilayers give some other results. At all temperatures a finite
longitudinal resistance is registered. It decreases exponentially
under lowering of temperature, but it does not vanish completely.
Moreover, the counterflow longitudinal resistivity
$\rho_{xx}^{CF}$ is even higher than the longitudinal resistivity
measured in the parallel current geometry $\rho_{xx}^{\parallel}$
(in which the electrical currents in the adjacent layers are equal
in value and flow in the same direction).

The excitons cannot give a contribution into the conductivity in
the parallel current geometry. Therefore, the parallel current
conductivity $\sigma_{xx}^{\parallel}$ should be much smaller than
the counterflow conductivity $\sigma_{xx}^{CF}$. The relation
between the longitudinal resistivities is another one. At zero or
small Hall conductivity the longitudinal resistivity is in inverse
proportion with the longitudinal conductivity. But for the
conductivity tensor with a large Hall component the longitudinal
resistivity is in direct proportion with the longitudinal
conductivity.  The first case is realized for the counterflow
geometry and $\rho_{xx}^{CF}\approx 1/\sigma_{xx}^{CF}$. The
parallel current geometry corresponds to the second situation and
$\rho_{xx}^\parallel\approx
\sigma_{xx}^\parallel/(\sigma_{xy}^\parallel)^2$.

Exciton superfluidity can be imperfect due to phase slips caused
by motion of vortices across the flow. In that case the
counterflow resistivity is nonzero and, in principle, it can be
larger than the parallel current resistivity. The smallness of
$\rho_{xx}^\parallel$  can be accounted, for instance, for a small
density of carriers that contribute into $\sigma_{xx}^\parallel$.

The specifics of the exciton superfluidity in bilayer quantum Hall
systems is that the vortices carry electrical charges. Such
vortices not only bring to a decay of superflow, but also give a
direct contribution into conductivity. As was argued in \cite{13},
due to such a two-fold role of vortices, the counterflow and
parallel current longitudinal resistivities coincide with each
other (in the absence of other essential factors that may
contribute to the conductivity).

The role of vortices in transport behavior of quantum Hall
bilayers was considered in Ref. \cite{14}. The consideration
\cite{14} is based on the coherence network model \cite{15}. In
this model superfluid excitons flow in a network formed by narrow
links, and the dissipation is caused by vortices that cross these
links. In \cite{14} the drag geometry experiments are analyzed. In
this geometry  \cite{8-1,9,10,11} the current flows through one
layer and the resistivities are measured in both layers. The
resistivities in the active (drive) layer
$\rho_{\alpha\beta}^{drive}$ and in the passive (drag) layer
$\rho_{\alpha\beta}^{drag}$ are connected with
$\rho_{\alpha\beta}^\parallel$ and $\rho_{\alpha\beta}^{CF}$ by
the relations
$\rho_{\alpha\beta}^{drive}=(\rho_{\alpha\beta}^\parallel
+\rho_{\alpha\beta}^{CF})/2$ and
$\rho_{\alpha\beta}^{drag}=(\rho_{\alpha\beta}^\parallel
-\rho_{\alpha\beta}^{CF})/2$, where $\rho_{\alpha\beta}^\parallel$
and $\rho_{\alpha\beta}^{CF}$ are the resistivities of the layer
that is used as an active one in the drag geometry (in case of
imbalanced bilayers $\rho_{\alpha\beta}^\parallel$ and
$\rho_{\alpha\beta}^{CF}$ are different for different layers).
 The resistivities $\rho_{xx}^{drive}$
and $\rho_{xx}^{CF}$ demonstrate thermally activated behavior. The
activation energy depends asymmetrically on the imbalance of
electron densities of the layers \cite{9,10}. As was shown in
\cite{14} this behavior can be described under assumption that
vortex mobilities are thermally activated quantities and their
activation energies are different for different specie of
vortices. The vortex configurations in strongly disordered quantum
Hall bilayers were studied in \cite{er}. It was shown that
disorder should be rather strong to provide vortex proliferation.

The ideas of \cite{13,14,15,er} are important for the
understanding of transport properties of quantum Hall bilayers.
Nevertheless, the question requires further study because a number
of essential features should be explained. First of all, in all
experiments the drag resistivity $\rho_{xx}^{drag}$  is negative
and the counterflow resistivity $\rho_{xx}^{CF}$ is larger than
the parallel current resistivity $\rho_{xx}^{\parallel}$.
Actually, it is one of key points because in case of perfect
exciton superfluidity $\rho_{xx}^{drag}$ should be positive and
equal to $\rho_{xx}^{drive}$. Then, the origin of the counterflow
Hall resistivity remains unclear. Also, the absolute value of
$\rho_{xx}^{CF}$ observed experimentally is quite large and one
can think about additional factors (beside vortices) that increase
$\rho_{xx}^{CF}$.

In this paper we present a model in which transport features of
quantum Hall bilayers are accounted for a special influence of
imperfectness. We assume that due to imperfectness the local total
filling factor deviates from unity and in some area $\nu_T<1$,
while in other areas $\nu_T>1$. In the areas of the first (second)
type positive-charged (negative-charged) vortices emerge. Besides,
we imply that in some defect areas the interlayer phase coherence
is destroyed completely. Using the effective medium approach
\cite{16,17,18} we compute the effective longitudinal and Hall
resistivities  for the counterflow, parallel current and drag
geometries. The results are in good agreement with the experiment.

\section{Vortices in exciton superfluids in quantum Hall bilayers}

There are four species of vortices in the system considered
\cite{5}. Below they are notated by the index $\kappa=1,2,3,4$.
The wave functions $|\kappa\rangle$ for the states with a single
vortex can be presented as
\begin{eqnarray}\label{2}
|1\rangle=\prod_{m=0}^M\left(u c_{m,1}^+ + v
c_{m+1,2}^+\right)|0\rangle, \cr |2\rangle=\prod_{m=0}^M\left(u
c_{m+1,1}^+ + v c_{m,2}^+\right)|0\rangle, \cr
|3\rangle=c^+_{0,2}\prod_{m=0}^M\left(u c_{m,1}^+ + v
c_{m+1,2}^+\right)|0\rangle,\cr
|4\rangle=c^+_{0,1}\prod_{m=0}^M\left(u c_{m+1,1}^+ + v
c_{m,2}^+\right)|0\rangle.
\end{eqnarray}
Here $c_{m,i}^+$ is the operator of creation for an electron in
the  layer $i$ in the state with definite angular momentum $m$:
$\Psi_m(r,\theta)=r^m e^{-i m\theta} \exp(-r^2/4 \ell^2)$, where
$\theta$ is the polar angle counted according to the right-hand
rule with respect to the magnetic field direction, and
$\ell=\sqrt{\hbar c/ e B}$ is the magnetic length. The $u$ and $v$
coefficients are $u=\sqrt{\nu}$ and $v=\sqrt{1-\nu}$, where
$\nu=\nu_1$ is the filling factor for the layer 1,
 and
$1-\nu=\nu_2$ is the filling factor for the layer 2, $\nu_i=2\pi
\ell^2 n_{e,i}$, and $n_{e,i}$ is the electron density in the
layer $i$.

Each vortex is characterized by distinct values of its vorticity
and electrical charge. The vortex charge $q_\kappa$ can be found
from the corresponding wave function (\ref{2}). One can see from
(\ref{2}) that the charge of a given vortex is fractional and is
concentrated in a certain layer. We notate this layer by
$i_\kappa$.

 The vortex is a state with a clockwise circular electrical
current in one layer and a counterclockwise current in the other
layer. We use the convention that the vortex has the positive
vorticity $\alpha_\kappa=+1$ if  the electrical current associated
with the vortex  is a counterclockwise in the layer 1 and
clockwise in the layer 2. The vortex parameters for each specie
are given in Table \ref{t1}.
\begin{table}
 \caption{The vorticity $\alpha_\kappa$, the electrical charge $q_\kappa$
 and the layer $i_\kappa$ where the charge is concentrated, for four species
 of vortices}\label{t1}
  \begin{center}
 \begin{tabular}{|c|c|c|c|}
  \hline
   $\kappa$ & $\alpha_\kappa$ & $q_\kappa$ & $i_\kappa$ \\
  \hline
  1 & -1 & $e(1-\nu)$ & 2 \\
  2 & +1 & $e\nu$ & 1 \\
  3 & -1 & $-e\nu$ & 2 \\
  4 & +1 & $-e(1-\nu)$ & 1 \\
  \hline
\end{tabular}

 \end{center}
\end{table}

The vortices nucleate in pairs. The components of the pair have
the opposite vorticities. In the absence of electron deficit or
excess the vortices in the pair have the opposite electrical
charges, as well. The Coulomb attraction results in an increase of
the binding energy of the pair. Therefore, free vortices at
$\nu_T=1$ emerge at temperatures higher than the
Berezinskii-Kosterlits-Thouless (BKT) transition
temperature\footnote{We do not consider the effect of vortex
unbinding caused by electical currents. It results in nonlinear
dependence of the voltage on the current.}

Deviation of the total filling factor from unity forces the
appearance of charge excitations that are transformed into vortex
pairs. In a given pair each vortex  has the fractional charge, but
the sum of the charges is integer. In areas with electron deficit
an equal number of vortices $\kappa=1$ and 2 is nucleated, and
equal number of vortices $\kappa=3$ and 4 emerge in areas with
electron excess. In a given area the vortices have the same sign
of charge and due to Coulomb repulsion the size of a bound vortex
pair can be quite large.

The energy of the pair is
\begin{equation}\label{17}
    E=\pi\rho_s \ln \frac{r}{a} + \frac{e^2 \nu(1-\nu)}{\varepsilon
    r},
\end{equation}
where $\rho_s$ is the superfluid stiffness, $r$ is the distance
between the vortices, and $a$ is the vortex core radius.  The
energy (\ref{17}) is minimum at the distance $r_b={e^2
\nu(1-\nu)}/{\pi\varepsilon\rho_s}$. The quantity $\gamma=r_b (2
n_{v})^{-1/2}$ (where $n_{v}$ is the concentration for one specie
of vortices) yields the ratio of the vortex pair size to the
average distance between the vortices. One can expect that at
$\gamma>1$ a plasma of free vortices of opposite vorticities
instead of a gas of vortex pairs emerges. Using the mean field
value of $\rho_s$ \cite{5} one finds that the condition $\gamma=1$
corresponds to the critical vortex density $n_{v,c}\approx 0.05
n_0$, where $n_0=1/2\pi \ell^2$ is the electron density of a
completely filled Landau level. The critical density $n_{v,c}$ is
proportional to $\rho_s^2$. It is known that the mean-field
approximation overestimates the superfluid stiffness $\rho_s$: it
does not take into account quantum and thermal excitations, and
the interaction with impurities that reduce the superfluid
stiffness (see, for instance, \cite{r1,r2, loz}). Therefore, the
actual critical density can be considerable smaller.

\section{The resistivity caused by vortex motion}

Let us consider the forces that act on a vortex. The vortices are
electrically charged and the electric and magnetic components of
the Lorentz force are applied to them: $${\bf F}_{L}=q_\kappa {\bf
E}_{i_\kappa}+ q_\kappa {\bf v}_\kappa \times {\bf B}/c,$$ where
${\bf E}_i$ is the electrical fields in the layer $i$ and ${\bf
v}_\kappa$ is the vortex velocity.

Vortices carry the vorticity and in system with nonzero net
current the Magnus force ${\bf F}_{M}$ emerges. This force can be
obtained from the derivative of the electrical current energy with
respect to the vortex position (see, for instance, \cite{abr}):
$${\bf F}_{M}=2\pi\alpha_\kappa\rho_s \nabla \varphi \times
\hat{z}.$$ Here $\hat{z}$ is the unit vector directed along the
magnetic field, and $\nabla \varphi$ is the gradient of the phase
of the order parameter taken far from the vortex center. The net
currents in the layers read as ${\bf j}_{s1}=-{\bf j}_{s2}=e
\rho_s \nabla\varphi/\hbar$.
 The Magnus force does not depend on the vortex velocity, because
in the case considered this force is proportional to the
difference of the net currents ${\bf j}_{s1}-{\bf j}_{s2}$. This
difference is the same in the lab reference frame and the
reference frame connected with a moving vortex.

We should  also take into account the viscous force ${\bf
F}_\eta=-\eta_\kappa {\bf v}_\kappa$, where $\eta_\kappa$ is the
viscosity parameter. In what follows we will imply the vortex
motion is connected with thermally activated hops of vortices
between pinning centers. Is this case temperature dependence of
the viscosity parameters can be approximated as
$\eta_\kappa\propto \exp(\Delta_\kappa/T)$

In the stationary state the resultant force applied to the vortex
is equal to zero
\begin{equation}\label{5}
2\pi\alpha_\kappa\rho_s \nabla \varphi\times\hat{z}-\eta_\kappa
{\bf v}_\kappa+q_\kappa {\bf E}_{i_\kappa}+ \frac{q_\kappa}{c}
{\bf v}_\kappa\times {\bf B}=0.
\end{equation}
Eq. (\ref{5}) is fulfilled for each specie $\kappa$ with nonzero
density.

The motion of vortices across the flow results in a decay of the
phase gradient. The rate of decay is $\partial
(\nabla\varphi)/\partial t =\sum_\kappa 2\pi n_\kappa
\alpha_\kappa {\bf v}_\kappa\times \hat{z}$, where $n_\kappa$ is
the density of vortices of specie $\kappa$. Antiparallel
electrical fields applied to the layers may compensate this decay.
They increase the phase gradient at the rate $\partial
(\nabla\varphi)/\partial t =(e/\hbar)({\bf E}_1-{\bf E}_2)$. In
the stationary state the following condition should be satisfied
\begin{equation}\label{6}
\sum_\kappa 2\pi n_\kappa \alpha_\kappa {\bf v}_\kappa\times
\hat{z}+\frac{e}{\hbar}({\bf E}_1-{\bf E}_2)=0.
\end{equation}

Having the vortex density one solves Eqs. (\ref{5}) and (\ref{6})
and finds the contribution of excitons and direct contribution of
vortices into conductivity. To compute the conductivity tensor one
should also take into account the bare conductivity of Landau
level.

In the BCS-like state each electron is distributed between two
one-particle states with the same guiding center index and
distinct layer indices. The system responds on the resultant
electrical force ${\bf F}_E=-N_e(\nu_1 e{\bf E}_1+ \nu_2 e{\bf
E}_2)$ ($N_e$ is the total number of electrons) in the same manner
as a completely filled Landau level responds on electrical field.
In the latter case the electron gas moves as a whole with the
drift velocity ${\bf v}_0=c {\bf E}\times {\bf B}/B^2$. In case of
the bilayer system the expression for the drift velocity contains
the effective field ${\bf E}_{eff}=\nu_1 {\bf E}_1+ \nu_2 {\bf
E}_2$. The velocity ${\bf v}_0$ can be also found from the
condition that the electrical and magnetic components of the
Lorentz force applied to the whole system compensate each other,
as it takes place for a completely filled Landau level with
negligible small relaxation
\begin{equation}\label{7}
   \nu_1 {\bf E}_1+\nu_2 {\bf E}_2+\frac{1}{c}{\bf
 v}_0\times {\bf B}=0.
\end{equation}
The motion of an electron gas  with the velocity ${\bf v}_0$
yields the contribution into  the electrical currents $\delta{\bf
j}_i=-e n_0 \nu_i{\bf v}_0$.

Thus, the electrical currents in the layer read as
\begin{eqnarray}\label{8}
 {\bf j}_1=-\sigma_0\nu_1\left(\nu_1{\bf E}_1+\nu_2{\bf
E}_2\right)\times \hat{z}\cr+\frac{e}{\hbar}\rho_s
\nabla\varphi+q_2 n_2 {\bf v}_2+q_4 n_4 {\bf v}_4,\cr {\bf
j}_2=-\sigma_0\nu_2\left(\nu_1{\bf E}_1+\nu_2{\bf
E}_2\right)\times \hat{z}\cr-\frac{e}{\hbar}\rho_s
\nabla\varphi+q_1 n_1 {\bf v}_1+q_3 n_3 {\bf v}_3,
\end{eqnarray}
where $\sigma_0=e^2/2\pi\hbar$

Tuning the magnetic field one can always fulfill the condition
$\nu_T=1$ in average. But due to structural defects the local
$\nu_{T}$ can be larger than unity in some areas and smaller than
unity in other areas. For simplicity, we consider the system with
two statistically distributed areas of equal fraction with local
$\nu_{T1}>1$ and $\nu_{T2}<1$ ($(\nu_{T1}+\nu_{T2})/2=1$). In
these areas the  vortex densities for certain $\kappa$ are
nonzero:
  $n_{1}=n_{2}=n_v$ in $\nu_T<1$ areas, and
$n_{3}=n_{4}=n_v $ in $\nu_T>1$ areas
($n_v=n_0(\nu_{T1}-1)=n_0(1-\nu_{T2})$).

We start from the analysis of transport properties of balanced
bilayers ($\nu_1=\nu_2$). In this case the electrical charges of
the vortices are equal in modulus. It is reasonable to assume that
the viscosity coefficients are the same for all species of
vortices ($\eta_1=\eta_2=\eta_3=\eta_4=\eta$).

Solving Eqs.(\ref{5}) and (\ref{6}) and substituting the solition
into Eq. (\ref{8}) we obtain
\begin{eqnarray}\label{9}
{\bf
j}_+=\frac{\sigma_0}{2}\Bigg[\frac{n_v}{n_0}\frac{\eta\beta}{\eta^2+\beta^2}{\bf
E}_+\cr-\left(1\pm\frac{n_v}{n_0}\frac{\beta^2}{\eta^2+\beta^2}\right){\bf
E}_+\times \hat{z}\Bigg],\cr {\bf
j}_-=\frac{\sigma_0}{2}\frac{n_0}{n_v}\left[\frac{\eta}{\beta}{\bf
E}_-\pm \left(1-\frac{2 n_v}{n_0}\right){\bf E}_-\times
\hat{z}\right],
\end{eqnarray}
In (\ref{9}) the upper(lower) sign corresponds to $\nu_{T}>1$
($\nu_{T}<1$) areas. Here we introduce the notation
$\beta=\pi\hbar n_0$, ${\bf j}_\pm={\bf j}_1\pm {\bf j}_2$ and
${\bf E}_\pm={\bf E}_ 1\pm{\bf E}_2$. The system considered is a
two-component isotropic conducting medium. Each component is
characterized by the parameters $\sigma_{s}^+=\sigma_{xx}^+$,
$\sigma_{a}^+=\sigma_{xy}^+$ (the parallel current conductivities)
and $\sigma_{s}^-=\sigma_{xx}^-$, $\sigma_{a}^-=\sigma_{xy}^-$
(the counterflow conductivities).

The exact expressions for the effective conductivities can be
obtained by the method developed in \cite{16,17}. In case of equal
fractions of the components the effective quantities read as
\begin{eqnarray}\label{29}
\langle\sigma_{xx}^\pm\rangle=\sqrt{\sigma_{s1}\sigma_{s2}}
\left(1+ \frac{(\sigma_{a1}-\sigma_{a1})^2}
{(\sigma_{s1}+\sigma_{s2})^2}\right)^{1/2},\cr
\langle\sigma_{xy}^\pm\rangle=\frac{\sigma_{a1}\sigma_{s2}
+\sigma_{a2}\sigma_{s1}} {\sigma_{s1}+\sigma_{s2}}.
\end{eqnarray}
Here  $\sigma_{s(a)n}$ stands for $\sigma_{s(a)n}^+$ in the
expression for $\langle\sigma_{\alpha\beta}^+\rangle$ and for
$\sigma_{s(a)n}^-$ in the expression for
$\langle\sigma_{\alpha\beta}^-\rangle$, the index $n=1,2$
numerates the components.

Using Eqs. (\ref{9}) and (\ref{29}), we obtain
\begin{equation}\label{36}
\langle\sigma_{xx}^{+}\rangle=\frac{\sigma_0}{2}\frac{n_v}{n_0}
\frac{\beta}{\sqrt{\eta^2+\beta^2}},\quad
\langle\sigma_{xy}^{+}\rangle=-\frac{\sigma_0}{2},
\end{equation}

\begin{equation}\label{37}
\langle\sigma_{xx}^{-}\rangle=\frac{\sigma_0}{2}\frac{n_0}{n_v}
\frac{\sqrt{\eta^2+\beta^2\left(1-\frac{2
n_v}{n_0}\right)^2}}{\beta},\ \langle\sigma_{xy}^{-}\rangle=0.
\end{equation}

The effective parallel current  and counterflow resistivities are
obtained from (\ref{36}),(\ref{37}) by the operation of inversion
of the conductivity tensor. We take into account that $n_v/n_0$ is
the small parameter. In the leading order in $n_v/n_0$ the
resistivities equal to
\begin{eqnarray}\label{11}
    \rho^{\parallel}_{xx}\approx\rho^{CF}_{xx}
    \approx\frac{2}{\sigma_0}\frac{n_v}{n_0}\frac{\beta}
    {\sqrt{\eta^2+\beta^2}},\cr
 \rho^{\parallel}_{xy}\approx \frac{2}{\sigma_0},\quad
 \rho^{CF}_{xy}=0,\cr
\rho^{drive}_{xx}\approx\frac{2}{\sigma_0}\frac{n_v}{n_0}\frac{\beta}
    {\sqrt{\eta^2+\beta^2}},\
\rho^{drag}_{xy}=\rho^{drive}_{xy}=\frac{1}{\sigma_0},\cr
\rho^{drag}_{xx}\approx -\frac{2}{\sigma_0}
\left(\frac{n_v}{n_o}\right)^2
\frac{\beta^3}{(\eta^2+\beta^2)^{3/2}}.
\end{eqnarray}
The negative sign of $\rho^{drag}_{xx}$ means that  the induced
drag voltage is opposite to the voltage drop in the drive layer.
Note that our model predicts rather small $\rho^{drag}_{xx}$ - it
is quadratic in the small parameter $n_v/n_0$.

One can see that for large viscosity  $\eta\gg \beta$ the
resistivities $\rho^{\parallel}_{xx}\approx\rho^{CF}_{xx}\approx
\rho^{drive}_{xx} \propto \exp(-\Delta/T)$, in an qualitative
agreement with the experiment. The  Hall resistivities
$\rho^{\parallel}_{xy}$, $\rho^{drag}_{xy}$ and
$\rho^{drive}_{xy}$ are in the quantitative agreement with
experimental data.

Let us now consider the case of imbalanced bilayers. In imbalanced
systems the vortex charges differ not only in sign but in absolute
value. The viscosity is caused by the interaction of vortices with
the  pinning centers, so one can assume that the corresponding
activation energy depends on the vortex charge. According to Table
\ref{t1}, we introduce two viscocity parameters
$\eta_a=\eta_1=\eta_4$ and $\eta_b=\eta_2=\eta_3$. For simplicity,
we restrict the analysis to the case of large viscosities
$\eta_a,\eta_b\gg \beta$. In this case one can neglect the
difference of $\sigma_{ik}$ in $\nu_T>1$ and $\nu_T<1$ areas.
Then, the relations between the currents and the fields are found
to be
\begin{eqnarray}\label{9-1}
{\bf
j}_+\approx\frac{\sigma_0}{2}\Bigg[\frac{n_v}{n_0}\frac{2\beta}{\eta_a+\eta_b}{\bf
E}_+-{\bf E}_+\times \hat{z} \cr
+\frac{\eta_a-\eta_b}{\eta_a+\eta_b}{\bf E}_-\times \hat{z}
\Bigg],\cr {\bf
j}_-\approx\frac{\sigma_0}{2}\Bigg[\frac{n_0}{n_v}\frac{2\eta_a\eta_b}
{\beta(\eta_a+\eta_b)}{\bf E}_-  -(\nu_2-\nu_1)^2 {\bf E}_-\times
\hat{z}\cr+ \frac{\eta_a-\eta_b}{\eta_a+\eta_b}{\bf E}_+  \times
\hat{z}\Bigg].
\end{eqnarray}

The inverse relations in linear in $n_v/n_0$ order are
\begin{eqnarray}\label{109}
{\bf
E}_1\approx\frac{1}{\sigma_0}\left[\frac{n_v}{n_0}\frac{2\beta}{\eta_b}{\bf
j}_1+{\bf j}_1\times \hat{z}+{\bf j}_2\times \hat{z} \right],\cr
{\bf
E}_2\approx\frac{1}{\sigma_0}\left[\frac{n_v}{n_0}\frac{2\beta}{\eta_a}{\bf
j}_2+{\bf j}_2\times \hat{z}+{\bf j}_1\times \hat{z} \right].
\end{eqnarray}
It follows from (\ref{109}) that
\begin{eqnarray}\label{13}
    \rho_{1,xx}^{CF}\approx \rho_{1,xx}^{drive}\approx\frac{1 }{\sigma_0}
    \frac{n_v}{n_0}
    \frac{2 \beta}{\eta_b},
\cr
    \rho_{2,xx}^{CF}\approx \rho_{2,xx}^{drive}\approx\frac{1 }{\sigma_0}
    \frac{n_v}{n_0}
    \frac{2 \beta}{\eta_a},
\end{eqnarray}
\begin{equation}\label{110}
   \rho_{1,xy}^{drive}\approx \rho_{2,xy}^{drive}\approx
\rho_{xy}^{drag}\approx {\sigma_0}^{-1},\quad
\end{equation}
\begin{equation}\label{111}
\rho_{1,xy}^{CF}=\rho_{2,xy}^{CF}=0,\quad \rho_{xx}^{drag}=0.
\end{equation}
Here $\rho_{n,\alpha\beta}$ notifies the resistivity in $n$-th
layer.

One can see that  the longitudinal resistivities of different
layers are determined by different viscosities. Therefore, the
resistivities may demonstrate different temperature dependences.
It was observed experimentally that the activation energy for the
longitudinal resistivities is higher in the layer with larger
electron concentration. It corresponds to the grows of
$\Delta_\kappa$ under increase of the vortex charge $q_\kappa$.

In addition, we note that in the leading order the Hall
resistivities (\ref{110}) are not sensitive to the imbalance as it
was seen in experiments.

\section{Possible origin of longitudinal drag resistivity}

We have shown that  due to specific relation between the vortex
charge and the filling factor $\nu$ the resistivities
$\rho_{1,xx}^{\parallel}$ and $\rho_{1,xx}^{CF}$ are very close to
each other. They almost cancelled each other in the expression for
the longitudinal drag resistivity, and  the latter quantity (see
Eq. (\ref{11})) is very small, in difference with experimental
data that show considerable negative resistivity.

It is important to  note that in case of perfect exciton
superfluidity the longitudinal drag resistivity should be positive
and equal to the drive resistivity. Indeed, any normal
conductivity channel contributes into $\sigma_{xx}^\parallel$, so
the resistivity $\rho_{xx}^\parallel$ should be nonzero. In the
counterflow geometry the normal channel is shunted by superfluid
excitons and cannot result in nonzero $\rho_{xx}^{CF}$. It yields
$\rho_{xx}^{drag}=\rho_{xx}^{drive}=\rho_{xx}^\parallel/2$

In case of slightly imperfect exciton superfluidity the situation
might be the following. An additional normal conductivity channel
(not connected with free vortices) increases both
$\sigma_{xx}^\parallel$ and $\sigma_{xx}^{CF}$. It results in an
increase of $\rho_{xx}^\parallel$, but in a decrease
$\rho_{xx}^{CF}$, so, it gives positive $\rho_{xx}^{drag}$. In
particular, bound vortex pairs (which are charged are carry zero
vorticity) should work in this direction. Thus, the question on
the origin of negative longitudinal drag remains open.

In this section we will show that the negative drag can be
accounted for the presence of defect areas, where the interlayer
phase coherence is suppressed completely. For simplicity, we
consider the case of balanced bilayers.

On a qualitative level the impact of defect areas can be
understood as follows. In these areas the counterflow and parallel
current conductivities are equal each other and coincide (at
least, approximately) with the conductivities of a single layer
with the filling factor $\nu=1/2$. The longitudinal conductivity
of a single layer $\sigma_{xx}^{single}$ is much smaller than the
counterflow conductivity determined by the exciton channel, and in
the counterflow geometry the defect areas work as a sort of an
exclude volume. Thus, the effective conductivity
$\sigma_{xx}^{CF}$ decreases and the effective resistivity
$\rho_{xx}^{CF}$ increases. On the other hand, if the conductivity
$\sigma_{xx}^{single}$ is of the same order as the conductivity
determined by the direct contribution of charged vortices the
defect areas have only a small impact on the resistivity
$\rho_{xx}^{\parallel}$.

To describe the influence of defect areas quantitatively we use
the approach developed in \cite{17}. According to \cite{17}, in
general case of unequal fractions of the components the effective
longitudinal and Hall conductivities of a two-component systems
are given by the expressions
\begin{eqnarray}\label{129}
\langle\sigma_{xx}^{\pm}\rangle=\frac{\sigma_{s1}\sigma_{s2}
(1-\lambda^2)f[p,\lambda]}
{D},\cr
\langle\sigma_{xy}^\pm\rangle=\sigma_{a1}-\frac{(\sigma_{a1}-\sigma_{a2})
\sigma_{s1}\lambda(1-f^2[p,\lambda])} {D},
\end{eqnarray}
where
$$D=\lambda(1-f^2[p,\lambda])\sigma_{s1}
+(f^2[p,\lambda]-\lambda^2)\sigma_{s2},$$
\begin{eqnarray}\label{130}
    \lambda=\frac{1}{4 \sigma_{s1} \sigma_{s2}}
    \Bigg[\sqrt{(\sigma_{s1}+\sigma_{s2})^2
    +(\sigma_{a1}-\sigma_{a2})^2}\cr -\sqrt{(\sigma_{s1}-\sigma_{s2})^2
    +(\sigma_{a1}-\sigma_{a2})^2}\Bigg]^2,
\end{eqnarray}
$p$ is the fraction of the component 1, and
\begin{eqnarray}\label{131}
f[p,\lambda]=-\left(\frac{1}{2}-p\right)(1-\lambda)\cr +
\sqrt{\left(\frac{1}{2} -p\right)^2 (1-\lambda)^2+\lambda }\ .
\end{eqnarray}
Eq. (\ref{129}) was obtained by mapping of the system with nonzero
$\sigma_{xy}$ to the system with zero Hall conductivity. The
function $f[p,x]$ approximates the effective conductivity $\langle
\sigma \rangle$ of an isotropic two-dimensional two-component
system with zero Hall conductivity: $\langle \sigma
\rangle=\sigma_1 f[p,\sigma_2/\sigma_1]$, where $\sigma_{i}$ is
the longitudinal conductivity of the $i$-th component, and $p$ is
the fraction of the component 1. The explicit expression
(\ref{131}) was obtained in the random resistor network approach
\cite{18}.

The three-component system can be reduced to the two-component one
in the following way. The defect areas are considered as the
component 1. The rest areas with imperfect exciton superfluidity
are considered as the component 2 with the conductivity equal to
the effective conductivity of the system considered in the
previous section.

Expressing the conductivities in $\sigma_0$ units, we have
\begin{eqnarray}\label{201}
{\sigma}_{s1}^{+}={\sigma}_{s1}^{-}=
\frac{\sigma_{xx}^{single}}{\sigma_0}=\alpha_1,\
{\sigma}_{a1}^{+}={\sigma}_{a1}^{-} \approx \frac{1}{2}, \cr
\sigma_{s2}^{+}=
\frac{n_v}{n_0}\frac{\beta}{2\sqrt{\eta^2+\beta^2}}=\alpha_2, \
{\sigma}_{a2}^{+}= \frac{1}{2}\cr
\sigma_{s2}^{-}=\frac{1}{4\alpha_2},\ \sigma_{a2}^{-}=0
\end{eqnarray}
The model contains two conductivity parameters $\alpha_1$,
$\alpha_2$ (normally, $\alpha_2\lesssim \alpha_1\ll 1$) and the
parameter $p$ (the defect area fraction).

The resistivity tensor is obtained by inversion of the effective
conductivity tensor. The effective resistivities depend on $p$.
Typical dependences are shown in Fig. 1.

\begin{figure}
\begin{center}
\includegraphics[height=.3\textheight]{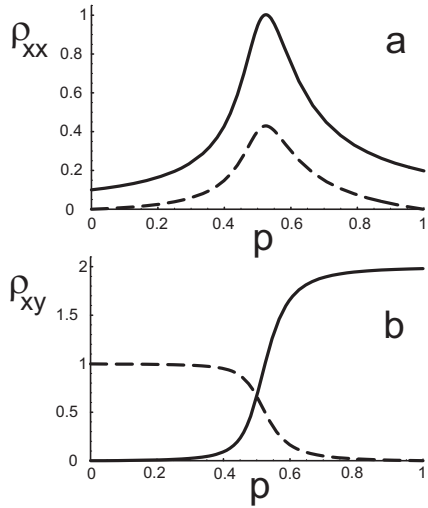}
\end{center}
  \caption{Longitudinal (a) and Hall resistivities (b)
  in units of $\sigma_0^{-1}$ versus
  the concentration of the areas with destroyed
  interlayer phase coherence. Solid curves - counterflow resistivities,
  dashed curves - drag resistivities.
  The longitudinal drag resistivity is given with reversed sign.
  The parameters $\alpha_1=0.05$
  and $\alpha_2=0.025$ are used.}
  \label{f3}
\end{figure}

One can see that the longitudinal drag resistivity is nonzero (and
negative). Other feature seen from Fig. 1 are the essential
increase of the longitudinal and the appearance of Hall
resistivity in the counterflow channel. The Hall drag resistivity
is equal approximately to $\sigma_0^{-1}$ (that correspond to the
experiment) if $p$ less and not to close to $1/2$.

\section{Conclusion}

Free vortices are dangerous for the superfluidity. But even in
two-dimensional superfluid systems nucleation of vortices does not
exclude perfect superfluidity. Below the BKT transition the
vortices of opposite vorticities bind in pairs. Motion of such
pairs, in difference with motion of single vortices, does not
result in phase slips. The question is why it is not the case for
quantum Hall bilayers.

In quantum Hall bilayers each elementary charge excitation
produces two fractionally charged vortices that repulse due to
Coulomb forces. There exists a critical charge excitation
concentration above which a gas of unpaired vortices emerges. Due
to renormalization of the superfluid stiffness the critical
concentration can be rather small. If the condition $\nu_T=1$ is
fulfilled in average and local total filling factor deviates from
unity, the excess electrons or holes induce vortex excitations
that suppress the superfluidity. If one trusts in the explanation
presented one could expect that concrete values of counterflow,
parallel current and drag resistivities are not universal
quantities and may vary from sample to sample.

The problem of negative longitudinal drag resistivity forces to
put forward the idea on the existence of areas with completely
destructed interlayer coherence.  This idea looks quite reasonable
taking into account that real geometry of arms should result in
essential edge effects (the presence of edge areas with destructed
interlayer coherence). The same effect can be caused by structure
defects and impurities in GaAs heterostuctures.

Nevertheless, one can hope that genuine exciton superfluidity can
be realized in bilayer quantum Hall systems. The only problem that
special requirements on purity and perfectness of the samples
should be fulfilled.


\begin{thebibliography}{99}
\bibitem{1}
S. I. Shevchenko, Sov. J. Low Temp. Phys. {\bf 2}, 251 (1976);
Phys. Rev. Lett. \textbf{72}, 3242 (1994); Phys. Rev. B {\bf 56},
10355 (1997).
\bibitem{1-1} Yu.
E. Lozovik, and V. I. Yudson,  Sov. Phys. JETP {\bf 44}, 389
(1976).
\bibitem{1-3} A. V. Balatsky, Y. N. Joglekar,
P.B. Littlewood, Phys. Rev. Lett. \textbf{93}, 266801 (2004); Y.
N. Joglekar, A. V. Balatsky, M. P. Lilly Phys. Rev. B \textbf{72},
205313 (2005).
\bibitem{1-5} E. Babaev, Phys. Rev. B \textbf{77}, 054512 (2008).
\bibitem{fer} H. A. Fertig, Phys. Rev. B \textbf{40}, 1087
(1989).
\bibitem{3} D. Yoshioka, A. H. MacDonald, J. Phys. Soc. Jpn.
\textbf{59}, 4211 (1990).
\bibitem{4} X. G. Wen and A. Zee, {Phys.
Rev. Lett.} \textbf{69}, 1811 (1992).
\bibitem{5} K. Moon, H. Mori,  K. Yang, S. M. Girvin, A. H. MacDonald,
L. Zheng, D. Yoshioka, and S. C. Zhang, {Phys. Rev.} \textbf{B51},
5138 (1995).
\bibitem{d1} S. M. Girvin, A. H. MacDonald, Multicomponent
quantum hall systems: the sum of their parts and more
Perspectives, in Quantum Hall Effects eds. S. D. Sarma and A.
Pinczuk, New York: Wiley, 1997.
\bibitem{d2} Yu. E. Lozovik and A. M. Ruvinsky, JETP \textbf{85}, 979
(1997).
\bibitem{d5} M. Abolfath, A. H. MacDonald, and L. Radzihovsky,
Phys. Rev. B\textbf{68}, 155318 (2003).
\bibitem{d6} A. I. Bezuglyj and S. I. Shevchenko, Phys. Rev. B
\textbf{75}, 75322 (2007).
\bibitem{5-1} S. H. Simon, Solid State Commun. {\bf 134}, 81
(2005).
\bibitem{8} M. Kellogg, J. P. Eisenstein, L. N. Pfeiffer, and K. W. West,
{ Phys. Rev. Lett.} {\bf 93}, 036801 (2004).
\bibitem{8-1} I. B. Spielman, M. Kellogg,
J. P. Eisenstein, L. N. Pfeiffer, and K. W. West, { Phys. Rev. B}
{\bf 70}, 081303 (2004).
\bibitem{9} R. D. Wiersma, J. G. S. Lok, S. Kraus, W. Dietsche, K. von
Klitzing, D. Schuh, M. Bichler,  H.-P.Tranitz, and W. Wegscheider
{ Phys. Rev. Lett.} {\bf 93}, 266805 (2004).
\bibitem{10} R. D. Wiersma, J. G. S. Lok, L. Tiemann, W. Dietsche,
K. von Klitzing, D. Schuh, W. Wegscheider, Physica E {\bf 35}, 320
(2006).
\bibitem{11} E. Tutuc, M. Shayegan, and D. A. Huse,  { Phys. Rev.
Lett.} {\bf 93}, 036802 (2004).
\bibitem{12}E. Tutuc, M. Shayegan, Phys. Rev. B  {\bf 72}, 081307(R)
(2005).
\bibitem{5-2} G. Moller, S. H. Simon, and E. H. Rezayi,
Phys. Rev. B {\bf 79}, 125106 (2009).
\bibitem{13} D. A. Huse, {Phys. Rev. B} {\bf 72}, 064514
(2005).
\bibitem{14} B. Roostaei, K. J. Mullen, H. A. Fertig, S. H. Simon,
Phys. Rev. Lett. {\bf 101}, 046804 (2008).
\bibitem{15} H. A. Fertig, and G. Murthy,
Phys. Rev. Lett. {\bf 95}, 156802 (2005).
\bibitem{er} P. R. Eastham, N. R. Cooper and D. K. K. Lee, Phys. Rev.
B \textbf{80}, 045302 (2009).

\bibitem{16} A. M. Dykhne, Sov. Phys. JETP {\bf 32}, 348 (1971).
\bibitem{17} D. Ya. Balagurov, Sov. Phys. JETP {\bf 81}, 1200 (1995).
\bibitem{18} S. Kirkpatrick, Rev. Mod. Phys. \textbf{45}, 574 (1973).
\bibitem{r1} Y. N. Joglekar and A. H. MacDonald, Phys. Rev. B {\bf 64},
155315 (2001).
\bibitem{r2} D. V. Fil, L. Yu. Kravchenko,
Low Temp. Phys. {\bf 35}, 712 (2009).
\bibitem{loz} O. L. Berman, Yu. E. Lozovik, D. W. Snoke and R.
D.Coalson, J. Phys.: Condens. Matter \textbf{19},  386219 (2007).
\bibitem{abr} A. A. Abrikosov, Fundamentals of the Theory of
Metals, North-Holland, Amsterdam, 1988.
\end{thebibliography}
\end{document}